\documentclass[twocolumn,preprintnumbers,elsart]{revtex4}
\usepackage{dcolumn}
\usepackage{bm}
\usepackage{graphicx}
\usepackage{amsmath}

\begin{document}

\title{Stable three-dimensional solitons in spin-orbit-coupled
atomic-molecular condensates}
\author{Chao Kong$^{1,2}$, Jinqing Li$^{1,2}$, Guihua Chen$^{3}$, Guilong Li$%
^{4}$, Zhaopin Chen$^{5}$, Haiming Deng$^{1,2}$\footnotemark[1], Boris A.
Malomed$^{6}$, Yongyao Li$^{7}$}
\footnotetext[1]{%
Corresponding author: woshidenghaiming@126.com }
\address{	\mbox{$^1$ School of Physics and Electronic-Electrical Engineering, Xiangnan University, Chenzhou 423000, China}\\
	$^2$ Microelectronics and Optoelectronics Technology Key Laboratory of Hunan Higher Education, Xiangnan University, Chenzhou 423000, China\\
	\mbox{$^3$ Department of Electronic Engineering, Dongguan University of Technology, Dongguan 523808, China} \\
	$^{4}$ College of Engineering and Applied Sciences, National Laboratory of Solid State Microstructures, Nanjing University, Nanjing 210023, China\\
	\mbox{$^{5} $ Physics Department and Solid-State Institute, Technion, Haifa 32000,Israel}\\
	%$^{6}$ Department of Physical Electronics, School of Electrical Engineering, Faculty of Engineering, Tel Aviv University, Tel Aviv 69978, %Israel\\
	\mbox{$^{6}$ Instituto de Alta Investigaci\'{o}n, Universidad de Tarapac\'{a},
	Casilla 7D, Arica, Chile}\\
	\mbox{$^{7}$ School of Physics and Optoelectronic Engineering, Foshan University, Foshan 5225, China}}

\begin{abstract}
We elaborate a mechanism for the creation of stable three-dimensional (3D)
solitons in spin-orbit-coupled (SOC) atomic-molecular Bose-Einstein
condensate, modeled by the mean-field equations with the quadratic
three-wave interaction, characterized by mismatch $\alpha $. The planar
(effectively two-dimensional) SOC is applied to the soliton's atomic
component, structuring it as a mixed mode (MM) or semi-vortex (SV). The
molecular component of the SV soliton is shaped as a 3D vortex, while the
molecular component in the MM soliton is an MM too. The solitons exist up to
a critical value of $\alpha $. The system demonstrates a relatively large
norm share of the vortex components, exceeding $50\%$ of the total norm,
which is an essential feature of SOC-supported solitons. This is a scheme for realizing stable vortex solitons in free space with the
quadratic nonlinearity.

\end{abstract}

\maketitle

Quadratic ($\chi ^{(2)}$) nonlinearity is a basic ingredient in many setups
of nonlinear optics and quantum matter (in particular, Bose-Einstein
condensates, BECs). In optics, $\chi ^{(2)}$ effects, such as the
second-harmonic generation, are well known in nonlinear crystals
%, such as lithium niobate (LiNbO$_{3}$), $\beta $-barium borate (BBO), and potassium
%titanyl phosphate (KTP)
\cite{Zumsteg1976,Chen1985,Fejer2001}. In BEC mixtures the quadratic
nonlinearity models the atomic-molecular conversion through photoassociation
\cite{Duine2004,Koehler2006}, which is a matter-wave counterpart of the $%
\chi ^{(2)}$ interaction in optics. The quadratic nonlinearities give rise
to many remarkable effects, such as collapse suppression and stabilization
of bright multidimensional solitons \cite{Torruellas19951,Torruellas1995},
which has made the $\chi ^{(2)}$ nonlinearity a basic means for the creation
of higher-dimensional solitons \cite%
{Rubenchik,Stegeman,HaoHe,Wise,Etrich2000,Buryak2002,Hayata1993,Corney2001,Lutsky2015,Saut,Bagci2021,B. Oles,P. D. Drummond,T. G. Vaughan}%
.

Multidimensional solitons exhibit rich phenomenology which is not available
in one dimension \cite{book}. In particular, the creation of vortex solitons
which carry the orbital angular momentum is a major objective \cite%
{Allen1992,Firth1997,Towers2001,Trapani2000,Hang2009,Adhikari2004,Baizakov2004,Leblond2007}%
. However, bright vortex solitons in quadratic nonlinear media face are
subject to the azimuthal instability, which leads to fragmentation of the
vortex structure \cite{Firth1997,LTorner,DVPetrov,Etrich2000,Buryak2002}.
This problem impedes the use of vortex solitons in applications to data
processing and optical manipulations. Various methods were proposed to
stabilize vortex solitons in quadratic nonlinear systems. A well-known
approach relies on the use of competing nonlinearities \cite%
{Xu2021,Xu2023,Mihalache2002,TJAlexander,DMihalache}. Thus, it has been
demonstrated that stable two-dimensional (2D) vortex solitons can be
maintained by the interplay of quadratic and repulsive cubic terms in
optical systems \cite{R. DeSalvo,DMihalache} and atomic-molecular BEC
mixtures \cite{TJAlexander}, similar to the stabilization of vortex solitons
by competing cubic-quintic nonlinearities \cite{EMWright,M. Quiroga,I.
Berezhiani,D. Mihalache,R. L. Pego,D. Mihalache2,C. Cartes,YYWang}.
Furthermore, stable quantum droplets in binary BEC can also carry vorticity,
due to the interplay of the cubic mean-field (MF) self-attraction and
beyond-MF quartic self-repulsion \cite%
{Barbut2016,Cabrera2018,Zhao2022,Luo2021,Yogurt2023,Nie2023,He2024,Kartashov2018,Li2018,Zhang2019}%
.

Without the resort to the repulsive cubic nonlinearity, the stabilization of
vortex solitons in quadratic nonlinear systems remains a challenging
problem. Several ways were proposed to address it, including the generation
of dissipative vortex solitons by introducing ring-shaped gain in lossy
media \cite{VELobanov}, and the stabilization of vortex solitons by means of
quasi-phase matching in nonlinear photonic crystals \cite{F. Zhao,C. Kong}.
However, the stability of vortex solitons in quadratic nonlinear systems
with the uniform background remains an open question.

To address the problem, we here propose a setup based on the
atomic-molecular BEC with the spin-orbit coupling (SOC) acting on the binary
atomic component. We demonstrate that this scheme, with the planar
(effectively 2D) SOC, makes it possible to create stable 3D solitons of the
semi-vortex (SV) and mixed-mode (MM) types. The 3D setup is modeled by the
system of coupled Gross-Pitaevskii equations (GPEs) for the MF wave
functions of the atomic ($\phi _{1,2}$) and molecular ($\psi $) components,
which includes the planar SOC of the Rashba type. In the scaled form, the
GPE system is
\begin{eqnarray}
i\frac{\partial \phi _{1}}{\partial t} &=&-\frac{1}{2}\nabla ^{2}\phi
_{1}+\lambda \left( \frac{\partial }{\partial x}-i\frac{\partial }{\partial y%
}\right) \phi _{2}-\psi \phi _{2}^{\ast },  \label{+} \\
i\frac{\partial \phi _{2}}{\partial t} &=&-\frac{1}{2}\nabla ^{2}\phi
_{2}-\lambda \left( \frac{\partial }{\partial x}+i\frac{\partial }{\partial y%
}\right) \phi _{1}-\psi \phi _{1}^{\ast },  \label{-} \\
i\frac{\partial \psi }{\partial t} &=&-\frac{1}{4}\nabla ^{2}\psi -\phi
_{1}\phi _{2}-\alpha \psi ,  \label{++}
\end{eqnarray}%
where $\lambda $ is the SOC\ strength, $\ast $ stands for the complex
conjugate, and $\alpha $ is the mismatch of the atomic-molecular
interconversion. The system conserves the energy (Hamiltonian),

\begin{gather}
E=\int \left[ \frac{1}{4}\left( 2\left\vert \nabla \phi _{1}\right\vert
^{2}+2\left\vert \nabla \phi _{2}\right\vert ^{2}+\left\vert \nabla \psi
\right\vert ^{2}\right) \right.  \notag \\
+\lambda \phi _{1}^{\ast }\left( \frac{\partial }{\partial x}-i\frac{%
\partial }{\partial y}\right) \phi _{2}+\lambda \phi _{1}\left( \frac{%
\partial }{\partial x}+i\frac{\partial }{\partial y}\right) \phi _{2}^{\ast }
\notag \\
\left. -\phi _{1}^{\ast }\phi _{2}^{\ast }\psi -\phi _{1}\phi _{2}\psi
^{\ast }-\alpha \left\vert \psi \right\vert ^{2}\right] \mathrm{d}x\mathrm{d}%
y\mathrm{d}z,  \label{E}
\end{gather}%
the total norm (scaled number of atoms),

\begin{equation}
N=\int (|\phi _{1}(\mathbf{r})|^{2}+|\phi _{2}(\mathbf{r})|^{2}+2|\psi (%
\mathbf{r})|^{2})d\mathbf{r\equiv }N_{1}+N_{2}+N_{3}\mathbf{,}  \label{N}
\end{equation}%
where $N_{1,2}$ and $N_{3}$ are the norms of the two atomic and single
molecular components. The set of the system's control parameters is $%
(N,\alpha )$, while $\lambda =1$ is fixed by scaling.

The SOC system conserves the angular momentum too \cite{Hnizdo,Scripta},
which in the present case implies the conservation of its $z$-component $%
M_{z}$. To derive the expression for it, the SOC terms in Eqs. (\ref{+}) and
(\ref{-}) are rewritten in terms of the polar coordinates $\left( r,\theta
\right) $, in the $\left( x,y\right) $ plane:%
\begin{eqnarray}
\lambda \left( \frac{\partial }{\partial x}-i\frac{\partial }{\partial y}%
\right) \phi _{2} &\equiv &\lambda \exp \left( -i\theta \right) \left( \frac{%
\partial }{\partial r}-\frac{i}{r}\frac{\partial }{\partial \theta }\right)
\phi _{2},  \notag \\
\lambda \left( \frac{\partial }{\partial x}+i\frac{\partial }{\partial y}%
\right) \phi _{2} &\equiv &\lambda \exp \left( +i\theta \right) \left( \frac{%
\partial }{\partial r}+\frac{i}{r}\frac{\partial }{\partial \theta }\right)
\phi _{2},  \label{theta}
\end{eqnarray}%
Thus, the full system of Eqs. (\ref{+})-(\ref{++}) is invariant with respect
to the continuous transform,%
\begin{equation}
\theta \rightarrow \theta +\delta \theta ,\phi _{1}\rightarrow \phi _{1}\exp
\left( -\frac{i}{2}\delta \theta \right) ,\phi _{2}\rightarrow \phi _{2}\exp
\left( +\frac{i}{2}\delta \theta \right) ,  \label{transform}
\end{equation}%
while $\psi $ is not affected. The application of the Noether theorem \cite%
{Thompson} to the system produces the expression sought for:%
\begin{equation}
\begin{aligned} M_{z} = \int \int \Bigg\{ & \bigg[
\frac{i}{2}\sum_{j=1,2}\phi _{j}^{\ast} \left( y\frac{\partial }{\partial
x}-x\frac{\partial }{\partial y}\right) \phi _{j} \\ & + i\psi ^{\ast
}\left( y\frac{\partial }{\partial x}-x\frac{\partial }{\partial y}\right)
\psi \bigg] + \mathrm{c.c.} \\ & + \frac{1}{2}\left( \left\vert \phi
_{1}\right\vert ^{2}- \left\vert \phi _{2}\right\vert ^{2}\right) \Bigg\}
dxdy, \end{aligned}  \label{Mz}
\end{equation}%
where $\mathrm{c.c.}$ stands for the complex conjugate.

Two components of the atomic wavefunction, $\phi _{1}$ and $\phi _{2}$, may
differ by values of the atomic spin, similar to three-wave interactions in
optics, where two components of the fundamental-harmonic wave differ by
mutually orthogonal polarizations \cite{Etrich2000,Buryak2002}.In this
respect, the molecular wavefunction $\psi $ is the counterpart of the
second-harmonic wave in the optical system.

Equations (\ref{+})-(\ref{N}) are normalized using physical scales
determined by the underlying BEC system. The characteristic length, energy,
and time are the healing length $L_{0}=1/\sqrt{8\pi n_{0}a_{s}}$, MF energy $%
E_{0}=4\pi \hbar ^{2}a_{s}n_{0}/m$, and $t_{0}=\hbar /E_{0}$, where $n_{0}$
is the peak density, $a_{s}$ the \textit{s}-wave scattering length, and $m$
the atomic mass. Accordingly, the original coordinates $\mathbf{\tilde{r}}$,
time $\tilde{t}$, and wavefunctions, $\tilde{\phi}_{1,2}$ and $\tilde{\psi}$%
, are scaled as $\mathbf{r}=\mathbf{\tilde{r}}/L_{0}$, $t=\tilde{t}/t_{0}$,
and $\phi _{1,2}=\tilde{\phi}_{1,2}/\sqrt{n_{0}}$, $\psi =\tilde{\psi}/\sqrt{%
n_{0}}$. The scaled total atom number $N$, detuning $\alpha $, SOC strength $%
\lambda $, and energy $E$ are expressed in units of $n_{0}L_{0}^{3}$, $%
E_{0}/\hbar $, $1/(L_{0}t_{0})$, and $E_{0}$, respectively.

Note that, while the self-focusing cubic nonlinearity leads to the commonly
known effect in the form of the critical and supercritical collapse in the
2D and 3D systems, respectively, the quadratic nonlinearity may lead to the
critical collapse only in the formal 4D case, thus being able to support
stable 3D fundamental (zero-vorticity) solitons \cite{Rubenchik,HaoHe}.
However, as mentioned above, the splitting instability of vortex solitons
remains a major problem.

The SOC effect gives rise to absolutely stable ground-state atomic solitons
of the SV and MM types in the 2D system with the cubic self-attraction \cite%
{Sakaguchi}, and metastable solitons of the same types in 3D \cite{HPu}. In
the present case, SV solitons can be produced by means of the imaginary-time
(IT) integration method \cite{Bao,HuangC,ZhongR} applied to Eqs. (\ref{+})-(%
\ref{++}) with the following input, written in terms of the cylindrical
coordinates $\left( r,\theta ,z\right) $:
\begin{gather}
\phi _{1}(t=0)=A_{1}\exp (-\alpha _{1}r^{2}-\beta _{1}z^{2}),  \notag \\
\phi _{2}(t=0)=A_{2}r\exp (-\alpha _{2}r^{2}-\beta _{2}z^{2}+i\theta ),
\notag \\
\psi (t=0)=\phi _{1}\phi _{2},  \label{input1}
\end{gather}%
where $A_{1,2}$, $\alpha _{1,2}$ and $\beta _{1,2}$ are positive real
constants, and the molecular component of the ansatz is selected according
to the phase-matching assumption, so that components $\phi _{2}$ and $\psi $
carry the same vorticity $1$, while $\phi _{1}$ has no vorticity.
Accordingly, the wave functions of the SV states are looked for as
\begin{equation}
\phi _{1}=e^{-i\mu t}\Phi _{1}(r,z),\phi _{2}=e^{-i\mu t+i\theta }\Phi
_{2}(r,z),\psi =e^{-2i\mu t+i\theta }~\Psi (r,z),  \label{ansatz}
\end{equation}%
where the chemical potential of the atomic components takes values $\mu \leq
\mu _{\max }=-\lambda ^{2}/2$ \cite{Sakaguchi}, $\Phi _{1,2}(r,z)$ and $\Psi
(r,z)$\ being real functions satisfying the radial equations:%
\begin{eqnarray}
\mu \Phi _{1} &=&-\frac{1}{2}\nabla _{r,z}^{2}\Phi _{1}+\lambda \left(
\partial _{r}+\frac{1}{r}\right) \Phi _{2}-\Psi \Phi _{2},  \notag \\
\mu \Phi _{2} &=&-\frac{1}{2}\left( \nabla _{r,z}^{2}-\frac{1}{r^{2}}\right)
\Phi _{2}-\lambda \partial _{r}\Phi _{1}-\Psi \Phi _{1},  \notag \\
2\mu \Psi &=&-\frac{1}{4}\left( \nabla _{r,z}^{2}-\frac{1}{r^{2}}\right)
\Psi -\alpha \Psi - \Phi _{1}\Phi _{2}.  \label{PhiPsi}
\end{eqnarray}%
where $\nabla _{r,z}^{2}\equiv \partial _{r}^{2}+\partial _{z}^{2}+\frac{1}{r%
}\partial _{r}$. At $r\rightarrow 0$, $\Phi _{2}$ and $\Psi $ vanish $\sim r$%
, while $\Phi _{1}$ remains finite. At $r\rightarrow \infty $, the
asymptotic form of the SV soliton is a product of the exponentially decaying
factor and oscillatory ones \cite{Sakaguchi}, $r^{-1/2}\exp \left( -\sqrt{%
-2\mu -\lambda ^{2}}r\right) \cos \left( \lambda r+\delta \right) $, with a
phase constant $\delta $.

3D solitons of the MM type can be produced by the IT simulations with the
following input \cite{Sakaguchi}:
\begin{gather}
\phi _{1,2}(t=0)=  \notag \\
A_{1}\exp (-\alpha _{1}r^{2}-\beta _{1}z^{2})\mp A_{2}r\exp (\mp i\theta
-\alpha _{2}r^{2}-\beta _{2}z^{2}),  \notag \\
\psi (t=0)=\phi _{1}\phi _{2}.  \label{input2}
\end{gather}%
As suggested by its name, the MM mixes vorticities $0$ and $\pm 1$ in its
components, so that the total $z$-component (\ref{Mz}) of this state is
zero. The chemical potentials of its atomic and molecular components remain $%
\mu $ and $2\mu $, although, unlike the SV ansatz (\ref{ansatz}), it is not
possible to introduce one for MM solitons with explicitly separated
coordinates.

%Solutions to Eqs. (\ref{+}), (\ref{-}) and (\ref{++}) for solitons of the SV type, with the real chemical potential $(\mu<0)$, can be looked for as (cf. Ref. \cite{Sakaguchi})%
%\begin{eqnarray}
%\phi _{1} &=&\exp \left( -i\mu t \right)f_{1}\left( r^{2}\right) ,  \notag \\
%~\phi _{2} &=&\exp \left( -i\mu t+i \theta \right)
%rf_{2}\left( r^{2}\right), \notag \\
%\psi &=&\exp \left( -2i\mu t+i \theta \right)
%rf_{3}\left( r^{2}\right),   \label{ansatz1}
%\end{eqnarray}%
%where $f_{1,2,3}(r^2)$ are the real amplitude functions. In the ansatz, we have assumed that the chemical potential of the molecular component is twice that of the atomic components.

\begin{figure}[tph]
\center
\includegraphics[height=2in,width=2.7in]{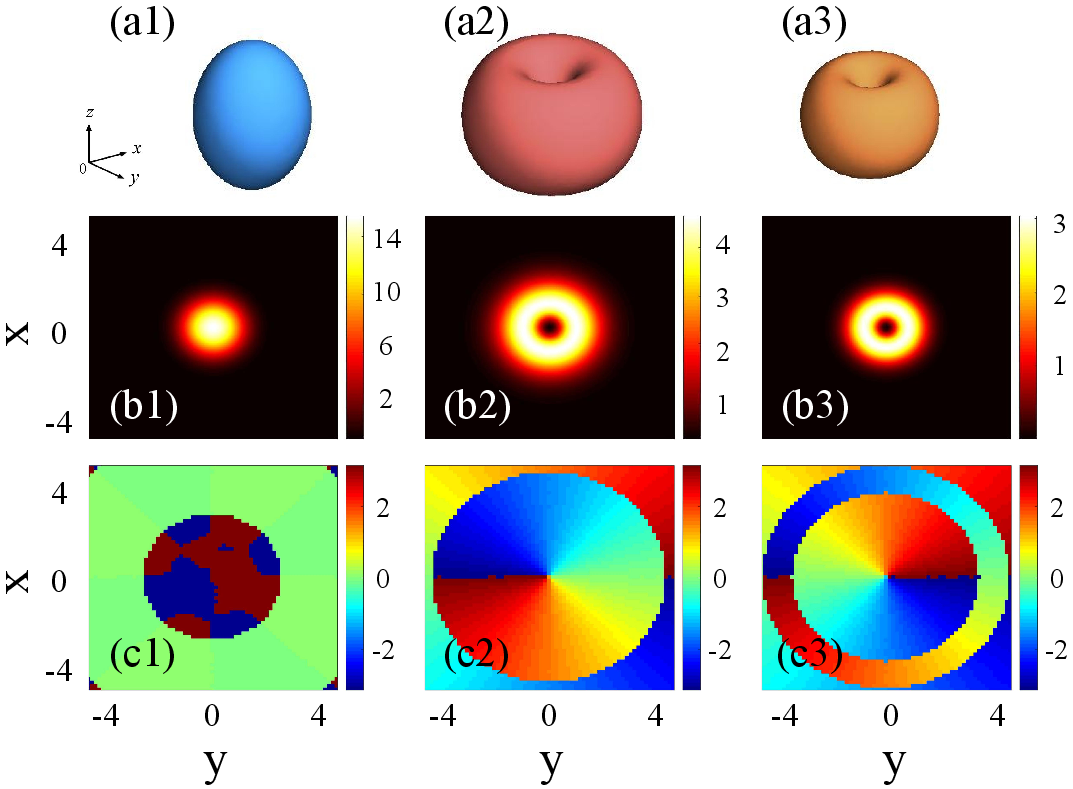}
\caption{A stable semi-vortex (SV) soliton with parameters $(N,\protect%
\alpha )=(200,0)$. Panels (a1), (b1), and (c1) display, severally, the
isosurface plot of the 3D density $\left\vert \protect\phi %
_{1}(x,y,z)\right\vert ^{2}$, 2D density distribution $\left\vert \protect%
\phi _{1}(x,y)\right\vert ^{2}$ in the $z=0$ plane, and the corresponding
phase pattern. Panels (a2), (b2), (c2) and (a3), (b3), (c3) display,
repsectively, the same for components $\protect\phi _{2}$ and $\protect\psi $%
. The angular momentum of this mode is $M_{z}$ $=66.2$.}
\label{fig1}
\end{figure}

Figure \ref{fig1} presents an example of a numerically found stable 3D SV
soliton, with parameters $(N,\alpha )=(200,0)$. In panels (a1), (a2), and
(a3) the isosurface plots of the atomic and molecular densities $\left\vert
\phi _{1,2}(x,y)\right\vert ^{2}$ and $\left\vert \psi (x,y)\right\vert ^{2}$%
, respectively, confirm the stability of the soliton after long-term
real-time propagation. The corresponding 2D cross-sections of the density
and phase in the $z=0$ plane, displayed alongside each 3D view, represent
the stationary solution produced by means of the IT-evolution method. The
characteristic SV structure is clearly seen in the figure: the atomic
component $\phi _{1}$ [Figs. \ref{fig1}(a1), (b1) and (c1)] features a
nearly isotropic density profile and a non-vortical phase pattern,
indicating zero topological charge, while the density of component $\phi _{2}
$ [Figs. \ref{fig1}(a2), (b2) and (c2)] exhibits a ring-shaped structure,
with the density vanishing at the center. The phase profile of $\phi _{2}$
shows the $2\pi $ circulation, confirming that it carries the unit
topological charge. The respective molecular field $\psi $ [Figs. \ref{fig1}%
(a3), (b3) and (c3)] inherits its topology from $\phi _{2}$, displaying a
ring density and the corresponding phase circulation.

\begin{figure}[tph]
\center
\includegraphics[height=2in,width=2.7in]{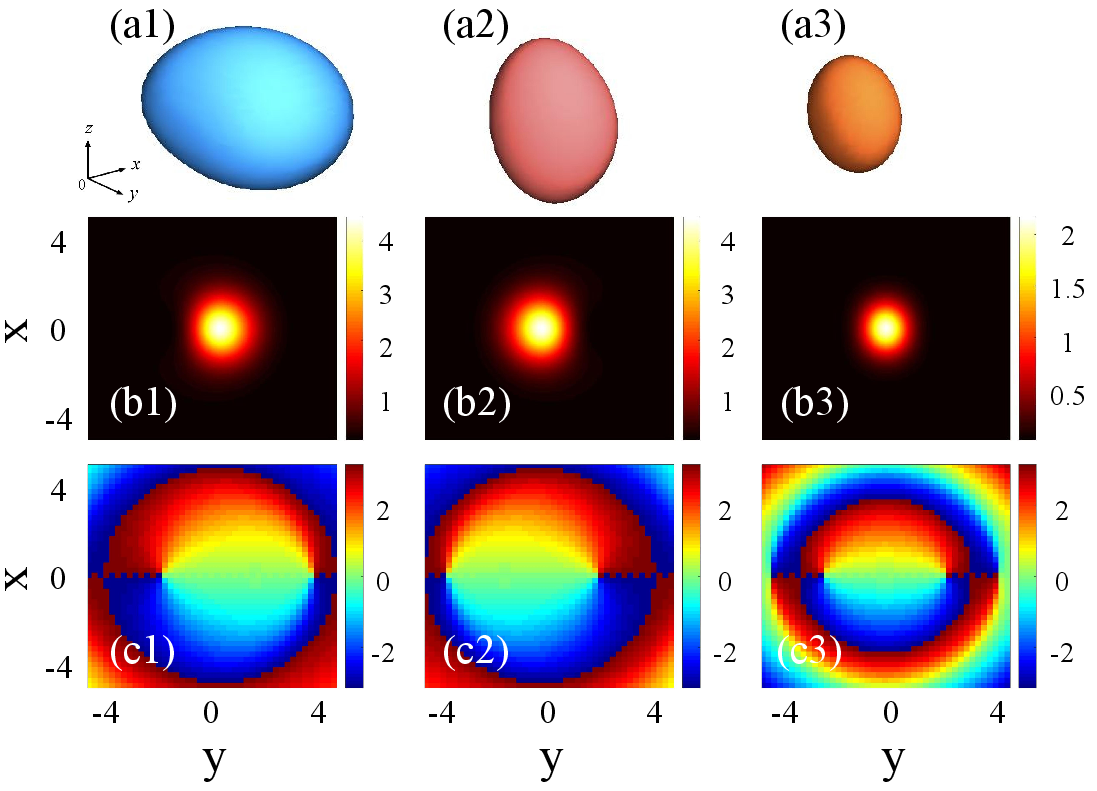}
\caption{A stable soliton of the MM type for $(N,\protect\alpha )=(100,0)$
and $M_{z}=0$. The meaning of panels is the same as in Fig. \protect\ref%
{fig1}.}
\label{fig2}
\end{figure}

An example of a stable MM soliton is showcased in Fig. \ref{fig2} for
parameters $(N,\alpha )=(100,0)$. The 3D isosurfaces in panels (a1-a3)
verify the robustness of the MM\ soliton in the real-time evolution, while
the 2D cross-sections illustrate the stationary profile. The hallmark of the
MM structure is the composite pattern of the atomic components, $\phi _{1}$
and $\phi _{2}$, each exhibiting the density profile in the cross-section
resulting from the superposition of the zero-vorticity mode and ones
carrying vorticities $\pm 1$. The molecular field $\psi $ exhibits a variety
of the MM pattern, with the dominant central density peak carrying a nearly
uniform phase, consistent with the expected cancellation of the vorticity in
the product of the atomic components.

\begin{figure}[tph]
\center
\includegraphics[height=1.25in,width=3.2in]{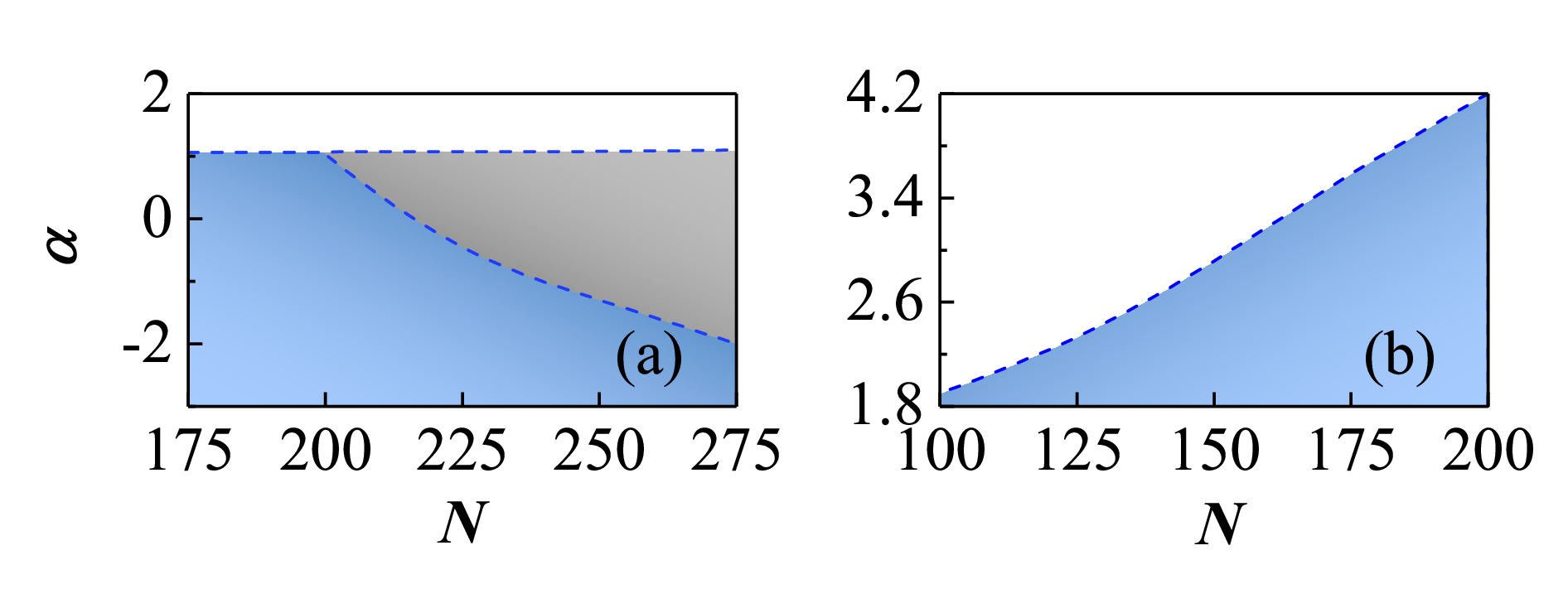}
\caption{Parameter regions of the SV and MM solitons in the $(N,\protect%
\alpha )$ plane. In (a), the SV-soliton region is includes stability (blue),
instability (gray), and non-existence (white) areas, with no solutions found
at $\protect\alpha >1.06$. In (b), the MM-soliton region features only
stability (blue) and non-existence (white) areas.}
\label{fig3}
\end{figure}

Parameter charts of the SV and MM solitons in the $(N,\alpha )$ plane are
plotted in Fig. \ref{fig3}. In panel (a), the SV solitons exist and are
stable or unstable in the blue and gray areas, respectively, and they do not
exist at $\alpha >1.06$. In panel (b) the MM solitons exist and are stable
in the blue region, and do not exist in the white one. The critical value of
the detuning, above which the MM solitons do not exist, gradually increases
with the growth of the particle number $N$.

\begin{figure}[tph]
\center
\includegraphics[height=2.78in,width=3.2in]{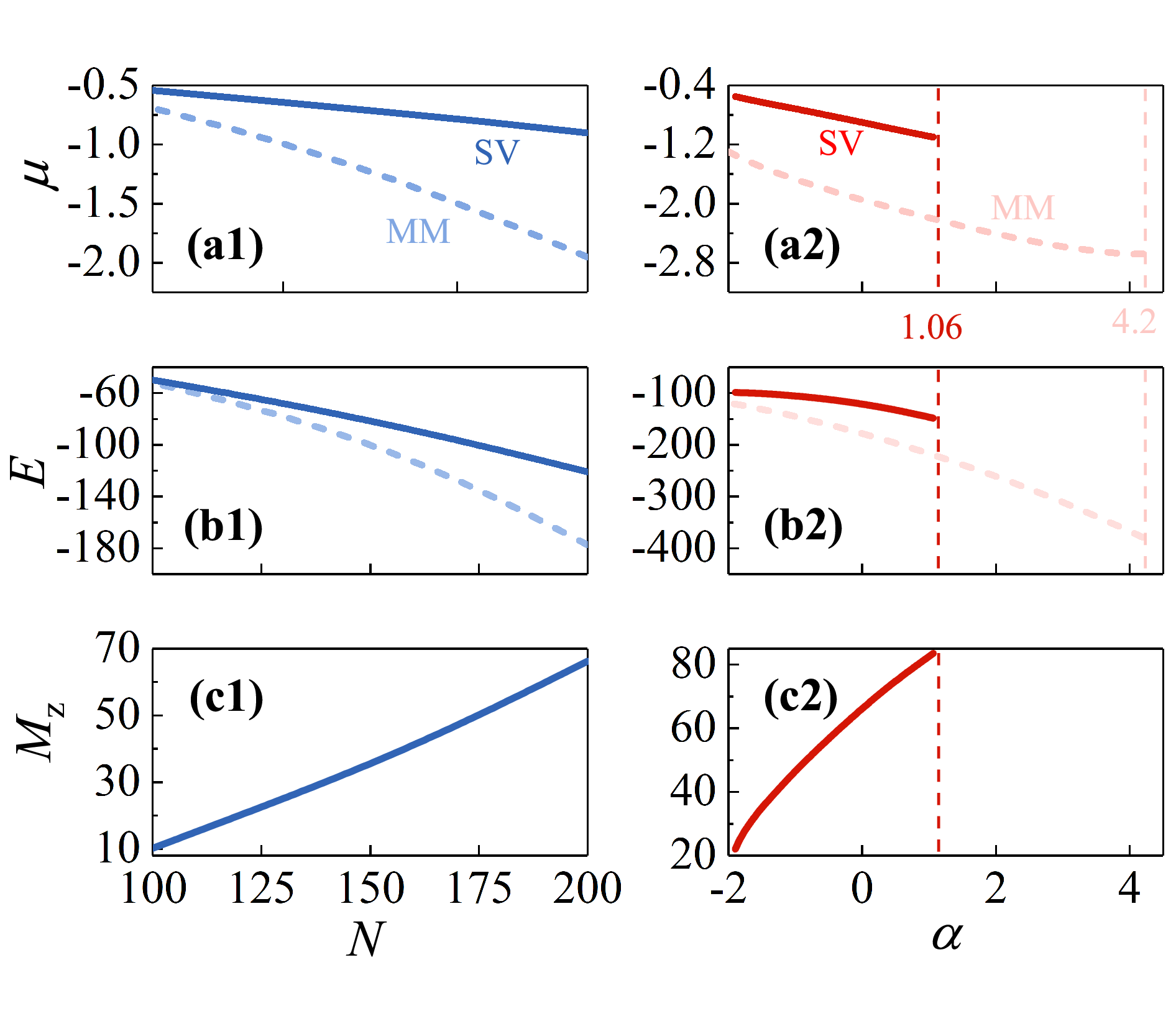}
\caption{The dependence of the soliton chemical potential $\protect\mu $,
energy $E$ and angular momentum $M_{z}$ on the particle number $N$ [panels
(a1-c1), with $\protect\alpha =0$], and on detuning $\protect\alpha $
[panels (a2-c2), with $N=200$]. The solid and dashed curves represent the 3D
SV and MM solitons, respectively. The solitons do not exists above the
critical value of the detuning, $\protect\alpha >\protect\alpha _{c}$.}
\label{fig4}
\end{figure}

Properties of the soliton families are further summarized in Fig. \ref{fig4}
by means of dependences of the chemical potential $\mu $, energy $E$, and
angular momentum $M_{z}$ on the atom number $N$ and detuning $\alpha $, with
the solid and dashed curves representing the SV and MM families,
respectively. In panels (a1)-(c1), where the detuning $\alpha =0$ is fixed,
both $\mu $ and $E$ decrease monotonously with the increase of $N$
increases. In particular, the negative slope $d\mu /dN<0$ is an indicator of
plausible stability of the solitons, as per the celebrated
Vakhitov-Kolokolov criterion \cite{VK,Berge,Kuz}. Further, a natural
property is that the angular momentum increases approximately linearly with $%
N$, as shown in Fig. 4(c1).

The dependences of $\mu $, $E$, and $M_{z}$ on detuning $\alpha $, with
fixed $N=200$, are plotted in Figs. \ref{fig4}(a2-c2). In this connection,
the numerical solution demonstrates that, at $\alpha <0$, the soliton's
spatial size grows with the increase of $|\alpha |$, asymptotically
approaching a spatially-uniform state, for which the chemical potential of
the atomic components attains it limit value, $\mu _{\max }=-\lambda ^{2}/2$%
. For $\alpha >0$, there exists a critical value, $\alpha _{c}$, above which
the SV and MM solitons do not exist.

To explain these findings, we employ the cascading-limit approximation \cite%
{Stegeman}, adiabatically eliminating the molecular field $\psi $ in the
case of strong detuning (large $-\alpha $), neglecting derivatives of $\psi $
in Eq. (\ref{++}): $\psi \approx -\phi _{1}\phi _{2}/\alpha $. Substituting
this approximation in Eqs. (\ref{+}) and (\ref{-}) yields simplified
equations,

\begin{eqnarray}
i\frac{\partial \phi _{1}}{\partial t} &=&-\frac{1}{2}\nabla ^{2}\phi
_{1}+\lambda \left( \frac{\partial }{\partial x}-i\frac{\partial }{\partial y%
}\right) \phi _{2}+\alpha ^{-1}\left\vert \phi _{2}\right\vert ^{2}\phi _{1},
\notag \\
i\frac{\partial \phi _{2}}{\partial t} &=&-\frac{1}{2}\nabla ^{2}\phi
_{2}-\lambda \left( \frac{\partial }{\partial x}+i\frac{\partial }{\partial y%
}\right) \phi _{1}+\alpha ^{-1}\left\vert \phi _{1}\right\vert ^{2}\phi
_{2}.\ \ \ \ \ \ \ \   \label{cascading}
\end{eqnarray}%
For $\alpha <0$, the cubic terms in Eqs. (\ref{cascading}) account for the
attractive interaction between the atomic components, promoting the
formation of solitons. In the case of $-\alpha \gg 1$, the SV soliton
gradually spreads out, turning into an indefinitely broad state with an
indefinitely small amplitude. For $\alpha >0$, the cubic terms mediate the
repulsive interaction, therefore, when positive $\alpha $ exceeds the
critical value $\alpha _{c}$, the repulsion overwhelms the SOC-induced
confinement, eliminaring the 3D solitons in the framework of the full system
of Eqs. (\ref{+})-(\ref{++}).
\begin{figure}[tph]
\center
\includegraphics[height=0.9in,width=3.2in]{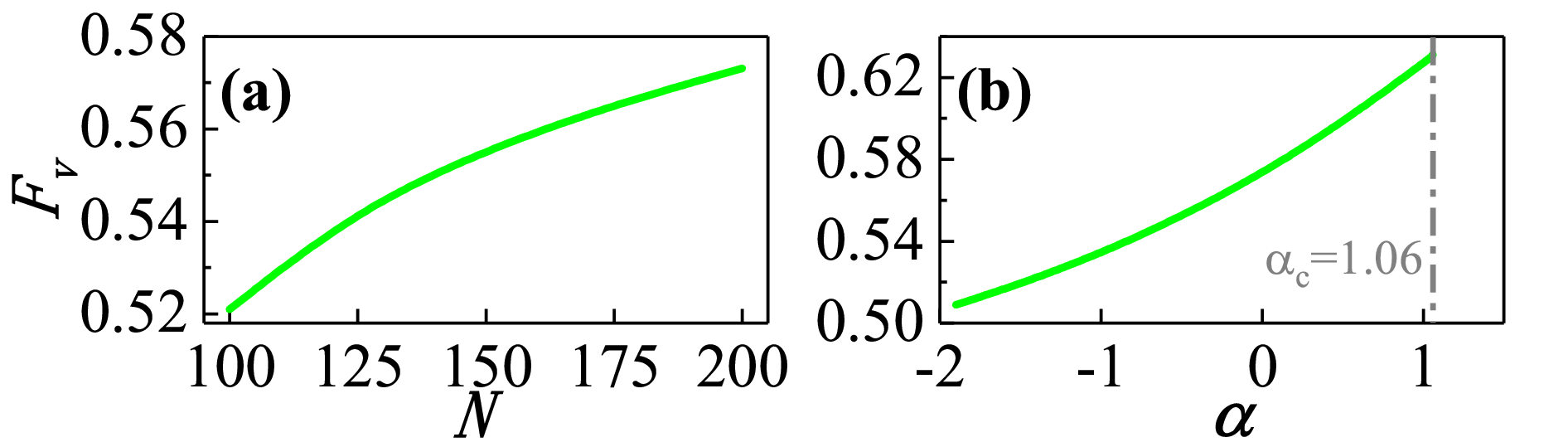}
\caption{The norm ratio of the vortex components in the SV solitons ($F_{v}$%
) versus $N$ (a) and $\protect\alpha $ (b). In (a) $\protect\alpha =0$ is
fixed, while in (b) $N=200$. }
\label{fig5}
\end{figure}

For the solitons of the SV type, vortex components are $\phi _{2}$ and $\psi
$, therefore for these states the norm share of the vorticity-carrying
components is defined as
\begin{equation}
F_{v}=\frac{N_{2}+N_{3}}{N},  \label{RF}
\end{equation}%
see Eq. (\ref{N}). To characterize the SV soliton family, in Figs. \ref{fig5}%
(a) and (b) we plot the norm share (\ref{RF}) vs. the total norm $N$ and
detuning $\alpha $. A key observation is that the vortex-norm share
consistently exceeds $50\%$ across the investigated parameter range. This is
an essential feature of SOC-supported solitons, as in the usual
atomic system the norm share of the vortex components is small \cite%
{Sakaguchi}.

For the experimental realization of the setup considered in this work, the
3D BEC can be prepared in a magnetic or optical trap by means of evaporative
cooling \cite{FDalfovo,M. H. Anderson}. The planar Rashba-type SOC can be induced,
using two Raman lasers, as demonstrated experimentally in Ref. \cite{JLin},
where the lasers couple the spin and momentum states of the atoms to induce
the SOC effect. Further, photoassociation can be used to couple atomic pairs
into the molecular bound state \cite{KMJones}. To connect with realistic
experimental conditions, we consider BEC in the ultracold gas of $^{39}$K
atoms \cite{JLin,CDM2021}, with the scattering length $a_{s}\approx 10a_{%
\mathrm{Bohr}}$ and peak density $n_{0}\approx 1.0\times 10^{14}\ \text{cm}%
^{-3}$, which produces the following scales: length $L_{0}\approx 0.866\
\mathrm{\mu }\text{m}$, time $t_{0}\approx 0.923\ \text{ms}$, and normalized
energy $E_{0}/h\approx 172\ \text{Hz}$. Based on this calibration, generic
parameters which occur in the above consideration imply the following
physical values: $N=100$ corresponds to $\simeq 6500$ atoms; detuning $%
\alpha =1$ corresponds to $\alpha _{\text{phys}}\simeq 100\ \text{Hz}$; the
evolution time $t=1000$ translates into $t_{\text{phys}}\approx 1\ \text{s}$%
; and energy $E=-100$ corresponds to $E_{\text{phys}}\simeq -1\times
10^{-29}\ \text{J}$, where $E<0$ implies a stable bound state of the system.

\emph{Conclusion }Our analysis reveals the existence of stable free-space 3D
solitons of the SV (semi-vortex) and MM (mixed-mode) types in the
atomic-molecular BECs system with the quadratic nonlinearity. The solitons
are stabilized by means of the SOC (spin-orbit-coupling) interaction,
applied solely to the atomic components. The planar SOC is sufficient for
the stabilization of the 3D solitons, while in the case of the binary atomic
system with the cubic self- and cross-attraction only the three-dimensional\
SOC can maintain the metastability of the SV and MM solitons in 3D \cite{HPu}%
. The molecular components of the SV and MM\ solitons feature the vortex and
MM\ structures, respectively. Basic characteristics of the solitons,
including the chemical potential, energy and angular momentum, are presented
as functions of the control parameters, \textit{viz}., the total norm and
mismatch of the atomic-molecular interconversion $\alpha $. The solitons do
not exist above the critical point, i.e., at $\alpha >\alpha _{c}>0$, which
is explained by means of the cascading approximation. In contrast to the
conventional SOC-BEC systems, in which the vortex components carry a small
share of the total norm, the atomic-molecular BEC supports stable 3D
solitons in which the share of the vortex parts is dominant, accounting for $%
>50\%$ of the total norm.

As a development of the present work, it is relevant to address mobility of
the solitons (the results of Ref. \cite{Sakaguchi} suggest that MM solitons
may be set in motion). It is relevant to construct stable \textquotedblleft
optical bullets" in the photonicl version of the system, based on the
three-wave parametric interaction and the photonic counterpart of SOC (cf.
Ref. \cite{Valery}). A challenging possibility is to consider excited stated
of the solitons, with extra vorticity added to both atomic components, and
the double vorticity added to the molecular one.

\begin{acknowledgments}
This work was supported by the Project of the National Natural Science
Foundation of China under Grants No. 12505031, No. 12675027, and No. 12274077, Hunan Provincial Natural
Science Foundation of China under Grant No.2024JJ5364, Guangdong Basic and Applied Basic Research Foundation under Grant No. 2024A1515010710, Scientific Research
Foundation of Xiangnan University for High-Level Talents, the Applied
Characteristic Disciplines of Electronic Science and Technology of Xiangnan
University, and Science and Technology Innovative Research Team in Higher
Educational Institutions of Hunan Province.
\end{acknowledgments}

\end{document}